\documentclass[12pt,preprint]{aastex}
\usepackage{amsmath}

\shorttitle{OH$^+$ in Diffuse Molecular Clouds}
\shortauthors{Porras et al.}

\begin{document}
\title{OH$^+$ in Diffuse Molecular Clouds}
\author{A.J. Porras\altaffilmark{1}$^,$\altaffilmark{2}, 
S.R. Federman\altaffilmark{1}, D.E. Welty\altaffilmark{3}, and 
A.M. Ritchey\altaffilmark{4}}
\altaffiltext{1}{Department of Physics and Astronomy, University of Toledo,
Toledo, OH 43606; steven.federman@utoledo.edu}
\altaffiltext{2}{Department of Physics and Astronomy, University of North 
Carolina, Chapel Hill, NC 27599; aporras@live.unc.edu}
\altaffiltext{3}{Department of Astronomy \& Astrophysics, University of 
Chicago, 5640 S. Ellis Ave., Chicago, IL 60637; dwelty@oddjob.uchicago.edu}
\altaffiltext{4}{Department of Astronomy, University of Washington, 
Seattle, WA 98195; aritchey@astro.washington.edu}

\begin{abstract}

Near ultraviolet observations of OH$^+$ and OH in diffuse molecular clouds 
reveal a preference for different environments.  
The dominant absorption feature in OH$^+$ 
arises from a main component seen in CH$^+$ (that with the highest 
CH$^+$/CH column density ratio), while OH follows CN absorption.  
This distinction provides new constraints on OH chemistry in these 
clouds.  Since CH$^+$ detections favor low-density gas with small fractions 
of molecular hydrogen, this must be true for OH$^+$ as well, confirming 
OH$^+$ and H$_2$O$^+$ observations with the $Herschel$ Space Telescope.  
Our observed correspondence indicates that the cosmic ray ionization 
rate derived from these measurements pertains to mainly atomic gas.  The 
association of OH absorption with gas rich in CN is attributed to the need 
for high enough density and molecular fraction before detectable amounts 
are seen.  Thus, while OH$^+$ leads to OH production, chemical arguments 
suggest that their abundances are controlled by different sets of conditions 
and that they coexist with different sets of observed species.  Of 
particular note is that non-thermal chemistry appears to play a limited role 
in the synthesis of OH in diffuse molecular clouds.

\end{abstract}

\keywords{ISM: lines and bands -- ISM: molecules -- ultraviolet: ISM -- 
astrochemistry}

\section{Introduction}

The chemistry of diffuse molecular clouds is mainly driven by 
ion-molecule reactions (e.g., van Dishoeck \& Black 1986; Le Petit et al. 
2004).  Many of the intermediate species are ions, and because they rapidly 
react with neutrals (especially H$_2$ or abundant atoms) and 
electrons, their abundances tend to be low.  A major pathway to OH involves 
the intermediates OH$^+$, H$_2$O$^+$, and H$_3$O$^+$ (e.g., Federman et al. 
1996 and references therein).  These oxygen-bearing 
ions are especially important; they provide a 
means to infer the cosmic ray ionization rate in diffuse molecular gas.  
This arises because cosmic rays ionize atomic and molecular hydrogen, 
leading to H$^+$ and H$_3^+$.  The hydrogen ions transfer their charge to O, 
producing first O$^+$ and then the molecular ions OH$_{\rm n}^+$ through 
ion-molecule reactions.

New, sensitive observations are now able to detect such intermediate 
species.  Wyrowski et al. (2010) observed the sub-mm
ground-state transitions of OH$^+$ 
in absorption from diffuse molecular clouds toward Sgr B2(M) 
with the Atacama Pathfinder Experiment.  Sub-mm measurements with 
the $Herschel$ Space Telescope revealed 
absorption from OH$^+$, H$_2$O$^+$, and H$_3$O$^+$ from diffuse molecular 
clouds toward distant star-forming regions (Neufeld et al. 2010; Gerin et al. 
2010; Ossenkopf et al. 2010).  Neufeld et al. showed that the observed 
OH$^+$/H$_2$O$^+$ ratio indicates that absorption from these molecules 
arises from material with small H$_2$ fractions.  The OH$^+$ 
abundance relative to H provided an estimate of the cosmic ray ionization 
rate, consistent with earlier determinations.  Absorption from OH$^+$ was 
also detected at near ultraviolet wavelengths through measurements acquired 
with the VLT/UVES instrument (Kre{\l}owski et al. 2010).  Kre{\l}owski et al. 
suggested that OH$^+$ absorption traces CH$^+$ the best among 
the molecular species seen at optical wavelengths.

We build on these earlier observations through a detailed analysis 
of OH$^+$ and OH absorption in the near ultraviolet.  While 
line-of-sight results are emphasized in earlier 
ground-based work, the focus here is on 
directions showing component structure with varying molecular abundances.  
We discuss results for VLT/UVES observations of BD$-$14$^{\circ}$5037, 
HD149404, HD154368, and HD183143 because they reveal the clearest 
separation between strong CH$^+$ and CN absorption.  Our previous work 
(e.g., Pan et al. 2005) showed that CH$^+$ absorption favors low density, 
molecule poor gas, while CN is associated with denser, molecule rich 
material.  This separation allows us to study the chemistry of 
oxygen-bearing species in diffuse molecular clouds in greater detail 
than was possible before.

\section{Observations}

Data below 4000 \AA, acquired with the Ultraviolet and Visual Echelle 
Spectrograph (UVES) of the Very Large Telescope (VLT) at Cerro Paranal, 
Chile, were obtained from the European Southern Observatory Science 
Archive Facility. Spectra of all four stars were acquired under 
program 065.I-0526 (PI: E. Roueff); additional data for HD149404 and 
HD154368 were obtained under program 082.C-0566 (PI: Y. Beletsky).  Both 
observing programs utilized the central wavelength setting 
at 3460 \AA\ and a 0.4 arcsec slit. The UVES pipeline reduction software was 
applied to the data in optimal extraction mode to produce merged spectra 
providing continuous coverage from 3050 to 3870 \AA.  
Finally, for each observing program, 
the individual exposures of a given target were coadded to produce a single 
high S/N spectrum. For HD149404 and HD154368, the spectral resolution is 
3.8 km s$^{-1}$ ($R=79,000$), since the observations employed CCD binning 
of $1\times1$. For BD$-14^{\circ}$5037 and HD183143, where $1\times2$ 
binning was used, the resolution is 4.2 km s$^{-1}$ ($R=71,000$).

Spectra of K~{\small I} $\lambda7699$ were also used in the 
analysis.  They came from archival UVES data (BD$-14^{\circ}$5037), Kitt 
Peak National Observatory Coude Feed data for HD149404 and HD154368 (Welty, 
unpublished), and McDonald Observatory 2.7 m coude observations for 
HD183143 (McCall et al. 2002).

Systematic fluctuations, with scale and amplitude similar to some of the 
true interstellar features and at very similar (geocentric) 
velocities, were noted near the wavelengths expected for the 
OH$^+$ line in the extracted spectra of several sight lines.
This apparent residual detector pattern is present even in the spectra of 
some stars with very weak absorption from other molecular species, where 
the OH$^+$ line would not be expected to be detectable.
The normalized spectra of those stars (HD116658, HD212571, HD182985) 
were used as templates to remove the residual detector pattern.  
Division by the template spectra generally yielded 
smoother continua and reduced noise near the OH$^+$ line.

The divided spectra were shifted to heliocentric velocities, combined 
(for sight lines observed under both programs), and normalized via 
low-order polynomial fits to the continuum regions.  For the four stars 
considered here, the S/N ratios measured in the continuum fits 
range from about 100--210 (per resolution element) near the OH lines 
($\lambda\lambda3078,3080$) to about 390--700 near the OH$^+$ line 
$\lambda3584$.  The resulting spectra, along with relatively weak 
lines of CH, CH$^+$, and CN, are shown in Figures 1 through 4.  When 
CN $\lambda3579$ was especially weak, the stronger line, $\lambda3874$, is 
displayed.

Total equivalent widths and column densities were obtained from the 
combined, normalized spectra by integrating the residual intensity and 
apparent optical depth (AOD), respectively, over the profiles of detected 
absorption lines.  The $1\sigma$ uncertainties on the equivalent widths 
include contributions from photon noise and continuum fitting, which 
are generally of comparable magnitude for relatively weak, narrow, 
unresolved lines.  Column densities and velocities for ``individual'' 
components discernible in the profiles of various atomic 
and molecular species were determined via independent multi-component fits 
to the profiles (e.g., Welty, Hobbs, \& Morton 2003).  Table 1 
summarizes the results for the molecular species; further details about 
the analysis on CH, CH$^+$, and CN will appear in 
Welty et al. (2013, in preparation, hereafter Welty2013).  For the weak OH and 
OH$^+$ lines, the total column densities obtained from the profile fits 
are consistent with the AOD estimates.

For the most part, the molecular data adopted in our analyses are the same 
as those used in previous studies (e.g., Federman et al. 1996; Felenbok \& 
Roueff 1996; Weselak et al. 2009a,b; Kre{\l}owski et al. 2012).  However, 
the wavelength for CH $\lambda3143$ and the OH$^+$ $f$-value are taken from 
more recent work.  The use of the wavelengths quoted by Lien (1984) for 
the CH $C-X$ band yielded components that were shifted by about 2 
km s$^{-1}$ relative to other lines of CH.  This discrepancy was removed 
when the data of Bembenek et al. (1997) were adopted.  As for the OH$^+$ 
$f$-value, the value quoted by de Almeida \& Singh (1981) and used by 
Kre{\l}owski et al. (2010) is no longer the preferred one.  The 
measurements of Brzozowski et al. (1974), which were utilized by 
de Almeida \& Singh (1981), were affected by space charge, resulting in 
shortened lifetimes.  These effects were analyzed by Curtis \& Erman 
(1977), who determined a corrected lifetime of $2.4\pm0.3$ $\mu$s; 
subsequent measurements by M\"{o}hlmann et al. (1978) based on an 
independent technique and theoretical calculations by Merch\'{a}n et al. 
(1991) yielded comparable lifetimes.  The recommended $f$-value for future 
studies is $1.14\times10^{-3}$, nearly a factor of 3 smaller than that 
considered in earlier studies.

\section{Results and Discussion}

Our total equivalent widths for CH, CH$^+$, CN, OH$^+$, and 
OH lines in the near ultraviolet are consistent with those of Weselak 
et al. (2009a, b) and Kre{\l}owski et al. (2010, 2012).  We note, 
however, that their values are often somewhat larger than ours, due perhaps 
to the additional data incorporated here and/or to slight differences in 
continuum placement.  The comparison is 
discussed in more detail in our larger survey of optical molecular 
absorption seen in UVES spectra (Welty2013).

Table 1 compiles the component structure for the four sight lines.  The 
column densities for CH, CH$^+$, and CN are based on all available data 
(Welty2013).  The OH features are associated with 
the strongest CN components.  
Absorption from OH$^+$, on the other hand, favors strong CH$^+$ components 
and especially those with the highest $N$(CH$^+$)/$N$(CH) ratio; 
these components tend to lack CN lines.  Kre{\l}owski 
et al. (2010) suggested that OH$^+$ is associated with CH$^+$, 
but their work focused on OH$^+$ detections.  Figures 1 to 4 illustrate the 
correspondences involving the strongest components in OH$^+$ and OH.  
Some possible weak OH$^+$ components (at $\le3\sigma$) may be related 
to weak components seen in strong lines of CH, CH$^+$, Na~{\small I} D, 
and/or K~{\small I} at longer wavelengths, which is representative of 
material dominated by atomic hydrogen.  The likelihood of these being 
detections is discussed below.  

As discussed in our previous work (e.g., Pan et al. 2005), different species 
have their largest abundances in specific regions within individual 
diffuse clouds.  Because CH$^+$ is destroyed by reactions with both 
atomic and molecular hydrogen, as well as electrons, it is 
generally found in low density gas with small H$_2$ fractions.  
Significant abundances of precursor molecules are required for observable 
amounts of CN, so that CN probes denser regions with larger H$_2$ 
fractions.  The CH radical is found in both types of gas, but larger column 
densities are usually seen in material traced by CN.

The same seems to apply to OH$^+$ and OH.  The connection between OH$^+$ and 
CH$^+$ (and possibly neutral atoms) indicates that OH$^+$ 
detections also favor low density gas.  Thus, our optical results 
are consistent with the conclusions of Neufeld et al. (2010) from analysis 
of sub-mm data acquired with $Herschel$.  In particular, 
absorption from OH$^+$ and H$_2$O$^+$ in diffuse molecular clouds occurs 
at velocities most similar to those seen in 21 cm absorption.  The inferred 
OH$^+$/H$_2$O$^+$ abundance ratios imply H$_2$ fractions less than about
10\%.  On the other hand, the association of OH with CN-rich gas reveals 
that OH is found in gas with higher densities, in excess of 
about 100 cm$^{-3}$ (e.g., Pan et al. 2005).

That difference in behavior is consistent 
with simple chemical arguments.  The 
intermediate species (OH$^+$) is rapidly destroyed by reactions with H$_2$, 
like the situation for CH$^+$.  The production of OH from OH$^+$ requires 
additional steps: reactions involving H$_2$ lead first to H$_2$O$^+$ and 
then H$_3$O$^+$, followed by dissociative recombination (H$_n$O$^+$ $+$ e).  
Denser gas also usually has greater amounts of extinction, which 
diminishes the effects of OH photodissociation.  This pictures runs 
counter to models that produce OH in gas heated by turbulent dissipation 
(e.g., Godard et al. 2009), heated by transient microstructure 
(Cecchi-Pestellini et al. 2009), or affected by the injection of hot H$_2$ 
(e.g., Cecchi-Pestellini et al. 2012) because OH is associated with CH$^+$ in 
these models.  Furthermore, a clearer picture emerges for the similarity 
in component structures seen in CN and CO spectra for directions with 
$N$(CO) $\ge$ few $\times10^{14}$ cm$^{-2}$ (e.g., Pan et al. 2005; 
Sonnentrucker et al. 2007; Sheffer et al. 2008).  
In such diffuse molecular clouds, CO arises from a series of reactions 
initiated by C$^+$ $+$ OH $\rightarrow$ CO$^+$ $+$ H, followed by 
production of HCO$^+$ and then CO through dissociative recombination.  
Thus, while OH$^+$ occasionally is seen in denser components, it is 
efficiently consumed there in the chain leading to OH and CO.

Instead, most of the production of OH, as well as all of the OH$^+$, 
therefore appears to be tied to cosmic ray ionization.  That behavior allows 
us to estimate the cosmic ray ionization rate in the components shown 
in Table 1.  Our analysis is based on the rate equations 
in Appendix of Federman et al. (1996).  While these expressions do not 
incorporate the effects of PAH species on ionization, Hollenbach et al. 
(2012) included these effects in their modeling, 
allowing us to estimate the changes 
that would occur.  We rely on rates and rate coefficients given in 
Federman et al. for the most part, but update the dissociative recombination 
coefficients and branching fractions 
when necessary from the compilation of Florescu-Mitchell \& 
Mitchell (2006).  The theoretical study of charge exchange between H$^+$ and 
O by Stancil et al. (1999) produces a rate coefficient very similar to the 
one used by us previously when a temperature of 80 K is considered.  A gas 
density of 100 cm$^{-3}$ is adopted here.  The complete list of inputs 
can be found in Welty2013.

The starting point is the steady-state rate equation for OH$^+$ in terms of 
fractional abundances, $x$(X) given by 

\begin{equation}
x({\rm OH}^+) = \frac{k_3 x({\rm H}_2)}{k_4 x({\rm H}_2)}
\frac{k_1 x({\rm O})}{k_2 x({\rm H}) + k_3 x({\rm H}_2)}
\frac{5 \zeta_p x({\rm H})}
{2 \alpha({\rm He}^+) x_{\rm e} n + k_1 x({\rm O}) n},
\end{equation}

\noindent where $k_i$ are rate coefficents, $\zeta_p$ is the primary cosmic 
ray ionization rate (equal to 0.5 $\zeta_{{\rm H}_2}$ given in Indriolo \& 
McCall 2012), $\alpha$(He$^+$) is the rate coefficient for He$^+$ 
recombination, $x_{\rm e}$ is the ionization fraction, and $n$ is the 
gas density [$n$(H) $+$ 2 $n$(H$_2$)].  For the gas rich in OH$^+$, we 
adopt $x$(H) = 10$x$(H$_2$).  The ionization fraction is taken to be 
$2\times10^{-4}$, so that it is slightly larger than only counting 
electrons from carbon ionization.  Solving for $\zeta_p$ with the rate 
coefficients noted above yields $\zeta_p$ $\approx$ 
$1.3\times10^{-8}$~$N$(OH$^+$)/$N$(H), where 
$N$(H) equals $N$(H~{\small I}) + 2$N$(H$_2$).  
For comparison, Hollenbach et al. 
(2012) provide a similar expression with a factor of $4.3\times10^{-8}$ 
for our adopted temperature and density.

The final step is the determination of $N$(H).  
For HD149404 and HD154368, values are available from Diplas \& Savage 
(1994) and Rachford et al. (2002).  For the 
other directions, we estimated $N$(H) from the total K~{\small I} 
column density, using the relationship in Welty \& Hobbs (2001).  We are 
confident in these estimates because the relationship gives values for 
$N$(H) toward HD149404 and HD154368 within about 10\% of the measured 
values.  Finally, we estimate $N$(H) for each component by determining 
its fraction of the total $N$(K~{\small I}) 
multiplied by the total $N$(H).  While 
this step is less secure, the relationship in Welty \& Hobbs spans the 
range for most of the values considered here.

For OH, we consider $x$(H) = $x$(H$_2$), $T$ = 50 K, and $n$ = 300 cm$^{-3}$, 
because its absorption appears to be associated with denser gas.  
Then eqns. (A2) and (A3) in Federman et al. (1996) are reduced to a simpler 
form for $x$(OH),

\begin{equation}
x({\rm OH}) = x({\rm OH}^+) \frac{k_4 x({\rm H}_2)}{k_6 x({\rm C}^+)}
\frac{\beta^{\prime}({\rm H}_3{\rm O}^+)}
{\beta({\rm H}_3{\rm O}^+)}.
\end{equation}

\noindent Here $\beta$(X) is the dissociative recombination rate coefficient 
for molecular ion X, with a prime denoting the branch leading to OH.  
In other words, once OH$^+$ is formed, OH arises directly from the sequence, 
OH$^+$ to H$_2$O$^+$ to H$_3$O$^+$ to OH without any branches for the 
adopted conditions.  OH destruction by reactions with C$^+$ is also 
more important than photodissociation for the sight lines studied here 
because the optical depth at ultraviolet wavelengths due to grains, 
$\tau_{\rm uv}\ge2$.  Combining our eqns. (1) and (2) and our 
set of input parameters leads to 
$\zeta_p$ $\approx$ $8.4\times10^{-10}$ $N$(OH)/$N$(H).

For most of the components shown in Table 1, our estimate for $\zeta_p$ 
ranges from 0.2 to $2.0\times10^{-16}$ s$^{-1}$ (see last two 
columns), consistent with other 
recent determinations.  For the one component with both OH$^+$ and OH, that 
toward BD$-14^{\circ}5037$ at about 5 km s$^{-1}$, the estimate from 
OH$^+$ is $0.5\times10^{-16}$ s$^{-1}$, while that from OH is 
$0.3\times10^{-16}$ s$^{-1}$.  They can be made to agree with slightly 
different values for $n$ (suggesting that the conditions vary along the 
line of sight because different conditions were adopted for the two 
estimates).  For that matter, all the estimates for a given sight line could 
be make consistent by slight changes of order a factor of a few in 
$n$.  Such relatively small changes are not unreasonable in light of earlier 
analyses of similar material.  Zsarg\'{o} \& Federman 
(2003) inferred densities of about 10 to a few hundred cm$^{-3}$ from neutral 
carbon excitation for directions with chemical conditions like those seen 
here for the OH$^+$ components.  Similarly, densities between about 100 and 
1000 cm$^{-3}$ seem appropriate for gas rich in CN (e.g., Sheffer et al. 
2008).  For instance, C$_2$ excitation for these clouds typically 
yields densities of 200 to 400 cm$^{-3}$ (e.g., Sonnentrucker 
et al. 2007).  Thus, these diffuse molecular clouds have primary 
cosmic ray ionization rates of about $1\times10^{-16}$ s$^{-1}$.  The 
exceptions to these results are the components only tentatively 
detected in OH$^+$ (and atomic species).  They suggest ionization 
rates greater than about $5\times10^{-16}$ s$^{-1}$ and are deemed suspect.

We can compare our estimates for $\zeta_p$ with others.  From analyses 
of H$_3^+$ absorption, Indriolo \& McCall (2012) found values ranging 
from $\le0.2$ to about $5\times10^{-16}$ s$^{-1}$, with the detections 
having uncertainties of 50\%.  In particular, Indriolo \& McCall 
obtained (in units of $10^{-16}$ s$^{-1}$) $\le0.2$ (BD$-14^{\circ}5037$), 
$\le1.3$ (HD149404), $2.10\pm1.31$ (HD154368), and $5.30\pm4.12$ and 
$3.91\pm2.96$ for the $-10$ and $+5$ km s$^{-1}$ components 
toward HD183143.  The general agreement is encouraging and 
could be made better by using the same set of physical conditions.  
Hollenbach et al. (2012) provide an analytic expression for the OH$^+$ 
abundance in diffuse molecular clouds that includes PAH species in the 
ionization balance.  For $T$ of 80 K, their relationship for $\zeta_p$ 
is $4.3\times10^{-8}$~$N$(OH$^+$)/$N$(H), with a proportionality constant 
about 3 times larger than ours.  This more comprehensive treatment 
suggests the need for us to start with densities more like 30 cm$^{-3}$ 
for the material with detectable amounts of OH$^+$ (and CH$^+$).

Fig. 3 in Hollenbach et al. shows chemical results for conditions 
appropriate for the clouds studied here.  
The OH abundance peaks at larger extinctions 
into a cloud, much like we infer.  The peak OH abundance relative to that 
for OH$^+$ is about 10, similar to our results for the $+5$ km s$^{-1}$ 
component toward BD$-14^{\circ}5037$.  The results in Table 1 reveal a 
large range in $N$(OH)/$N$(OH$^+$), however, from less than 3 to greater 
than 60.  More detailed comparisons require modeled column densities.

Indriolo et al. (2012) compared estimates for the cosmic ray ionization 
rate in diffuse clouds toward W51 from sub-mm observations of 
OH$^+$ and H$_2$O$^+$ with H$_3^+$ measurements in nearby sight lines.  
The analysis is based on the premise that OH$^+$ and H$_2$O$^+$ probe 
predominantly atomic gas, while H$_3^+$ is associated with H$_2$.  To 
bring the estimates into agreement, they deduced an efficiency factor 
for producing OH$^+$ from ionization of H atoms of about 0.10.  The 
comparisons between our results and those of Indriolo \& McCall (2012) do not 
require a low efficiency factor.  More precise data and detailed modeling 
are needed to clarify the situation.

\section{Conclusions}

VLT/UVES measurements reveal that absorption from OH$^+$ and OH arise 
from different environments.  Gas with lower densities and smaller H$_2$ 
fractions provide the conditions for observable amounts of OH$^+$; this 
material is also rich in CH$^+$.  Our results at near ultraviolet 
wavelengths confirm the conclusions reached by studying sub-mm absorption 
(e.g., Neufeld et al. 2010).  The OH components, on the other hand, are 
associated with CN-rich gas with higher densities and larger H$_2$ fractions.  

Since OH absorption is generally not seen in the strongest CH$^+$ components, 
the chemical routes for the two molecules must differ.  We suggest that 
cosmic ray ionization is the dominant source for OH in diffuse molecular 
clouds.  Simple chemical arguments are used to infer the cosmic ray 
ionization rate, finding values of $\approx$ $1\times10^{-16}$ s$^{-1}$.  
For the one component with both OH$^+$ and OH absorption in our sample, 
the analysis indicates that the conditions along the line of sight may 
vary somewhat, with temperatures varying about a factor 2 and densities about 
a factor of 3.  Such variation is not unreasonable, as it is seen in 
more comprehensive modeling efforts (e.g., van Dishoeck \& Black 1986).  Our 
estimates are also consistent with those inferred from H$_3^+$ 
observations (e.g., Indriolo \& McCall 2012) and from chemical models 
including the effects of PAH molecules (Hollenbach et al. 2012).  
The differences can easily be accommodated with small changes in 
gas density.  These findings will be incorporated 
into our survey (Welty2013) of a large number of 
directions to discern the range of densities and cosmic ray ionization 
rates for diffuse molecular clouds.

\acknowledgments
This work was supported in part by NASA grant NNG06GC70G (S.R.F.).  
A.J.P. acknowledges support by the National Science Foundation under 
Grant No. 0353899, and D.E.W. by AST-1238926.

\begin{deluxetable}{lrcrrcrrcrrcrrcrcc}
\tabletypesize{\scriptsize}
\rotate
\tablecolumns{18}
\tablecaption{Component Structures \label{tab:comps}}
\tablewidth{0pt}

\tablehead{
\multicolumn{1}{l}{Star}&
\multicolumn{3}{c}{CH}&
\multicolumn{3}{c}{CH$^+$}&
\multicolumn{3}{c}{CN}&
\multicolumn{3}{c}{OH$^+$}&
\multicolumn{3}{c}{OH}&
\multicolumn{2}{c}{$\zeta_p$}\\
\multicolumn{1}{c}{ }&
\multicolumn{1}{c}{$N12$}&
\multicolumn{1}{c}{$b$}&
\multicolumn{1}{c}{$v_{\odot}$}&
\multicolumn{1}{c}{$N12$}&
\multicolumn{1}{c}{$b$}&
\multicolumn{1}{c}{$v_{\odot}$}&
\multicolumn{1}{c}{$N12$}&
\multicolumn{1}{c}{$b$}&
\multicolumn{1}{c}{$v_{\odot}$}&
\multicolumn{1}{c}{$N12$}&
\multicolumn{1}{c}{$b$}&
\multicolumn{1}{c}{$v_{\odot}$}&
\multicolumn{1}{c}{$N13$}&
\multicolumn{1}{c}{$b$}&
\multicolumn{1}{c}{$v_{\odot}$}&
\multicolumn{2}{c}{ }}

\startdata
BD$-$14~5037 &  3.0 & [2.0] & $-$13.2 &  3.9 & [3.6] & $-$12.8 &      &       & \nodata &  2.7 & [2.0] & $-$11.9 & $\le 2.8$ &       & \nodata & 1.6 & \nodata\\
             & 36.0 &  1.1  &  $-$6.8 &  9.7 & [2.5] &  $-$7.4 & 19.5 & [0.6] &  $-$6.7 & $\le 2.9$ &       & \nodata & 12.4 & [2.5] &  $-$5.8 & \nodata & 0.4\\
             &      &       & \nodata & 18.0 & [2.5] &     1.0 &      &       & \nodata &      &       & \nodata &      &    & \nodata & \nodata & \nodata\\
            & 45.0 &  2.5  &     3.3 & 36.2 & [2.5] &     5.4 &  6.5 & [0.8] &     3.2 &  8.9 &  3.5  &     4.7 &  7.8 & [2.5] &     4.3 & 0.5 & 0.3\\
             & 10.0 & [2.5] &     8.7 &      &       & \nodata &  2.1 & [0.8] &     9.0 &      &       & \nodata &      &    & \nodata & \nodata & \nodata\\
 & \\
HD 149404    &  6.2 &  2.0  & $-$18.7 &  5.3 &  2.8  & $-$16.9 &  0.6 & [1.0] & $-$19.6 &  1.9 & [2.5] & $-$19.2 & $\le 1.7$ &       & \nodata & 0.2 & \nodata\\
             &      &       & \nodata &      &       & \nodata &      &       & \nodata &  4.0 & [2.5] & $-$12.8 & $\le 1.7$ &       & \nodata & 5.5 & \nodata\\
             &  6.3 &  2.1  &  $-$7.8 & 18.9 &  3.3  &  $-$6.7 &      &       & \nodata &  5.0 & [1.0] &  $-$7.0 & $\le 1.6$ &       & \nodata & 1.2 & \nodata\\
             &  2.9 & [1.2] &  $-$3.5 &      &       & \nodata &      &       & \nodata &  3.0 & [2.0] &  $-$1.8 & $\le 2.5$ &       & \nodata & 1.3 & \nodata\\
             & 12.4 &  1.1  &     0.6 & 22.0 &  1.9  &     0.0 &  2.9 & [1.0] &  $-$0.4 & $\le 3.0$ &       & \nodata &  4.3 &  2.5  &     1.1 & \nodata & 0.3\\
 & \\
HD 154368    &  2.1 & [1.0] & $-$20.8 &  0.5 & [2.5] & $-$22.1 &      &       & \nodata &      &       & \nodata &      &    & \nodata & \nodata & \nodata\\
             &      &       & \nodata &  1.1 & [2.5] & $-$14.8 &      &       & \nodata &      &       & \nodata &      &    & \nodata & \nodata & \nodata\\
             &      &       & \nodata &      &       & \nodata &      &       & \nodata &  1.7 & [1.5] & $-$10.7 & $\le 2.0$ &      & \nodata & 3.3 & \nodata\\
             &  9.4 & [1.2] &  $-$5.4 & 13.7 & [2.0] &  $-$5.6 &      &       & \nodata &  4.2 & [2.0] &  $-$5.1 & $\le 2.4$ &       & \nodata & 0.5 & \nodata\\
             & 47.5 &  1.1  &  $-$2.4 &  7.5 & [2.0] &  $-$1.5 & 24.3 &  0.8  &  $-$2.5 & $\le 2.9$ &       & \nodata & 17.9 &  1.5  &  $-$2.0 & \nodata & 0.8\\
             &  2.1 & [1.0] &     1.3 &      &       & \nodata &      &       & \nodata &  1.6 & [1.5] &     0.5 & $\le 2.6$ &       & \nodata & 1.3 & \nodata\\
             &      &       & \nodata &  0.4 & [2.5] &     3.5 &      &       & \nodata &      &       & \nodata &      &    & \nodata & \nodata & \nodata\\
 & \\
HD 183143    & 18.3 &  3.1  &  $-$9.5 & 31.5 &  2.4  &  $-$10.6 & 0.5 & [1.0] & $-$10.2 & 11.0 &  1.1  & $-$11.6 & $\le 3.9$ &       & \nodata & 1.9 & \nodata\\
             &      &       & \nodata &      &       & \nodata &      &       & \nodata &  2.8 & [1.5] &  $-$4.9 & $\le 4.0$ &       & \nodata & 6.1 & \nodata\\
            & 2.3 & [1.5] & 0.5 & 3.9 & [2.0] & 0.6 &      &       & \nodata &  3.1 & [1.5] &     1.9 & $\le 5.1$ &       & \nodata & 1.6 & \nodata\\
             & 30.9 &  2.0  &     5.3 & 37.4 &  1.8  &     5.3 &  3.2 & [1.0] &     5.7 & $\le 3.6$ &       & \nodata &  7.7 & [2.5] &     5.6 & \nodata & 0.6\\
\enddata
\tablecomments{Column densities (in cm$^{-2}$) are multplied by power of ten 
given in header; $b$ and $v_{\odot}$ are in km s$^{-1}$.  OH$^+$ and OH 
upper limits are based on fits with $b$ and $v_{\odot}$ taken 
from results for detected species in the component.  The cosmic ray 
ionization rates have units of $10^{-16}$ s$^{-1}$.  
Values for $b$ in square braces were fixed in profile fits.  The sum of the 
results for $J=0$ and $J=1$ are shown for the CN column density.  
The first (second) entry for $\zeta_p$ comes from our OH$^+$ (OH) 
analysis}
\end{deluxetable}

\clearpage

\begin{figure}
\begin{center}
\includegraphics[scale=0.8]{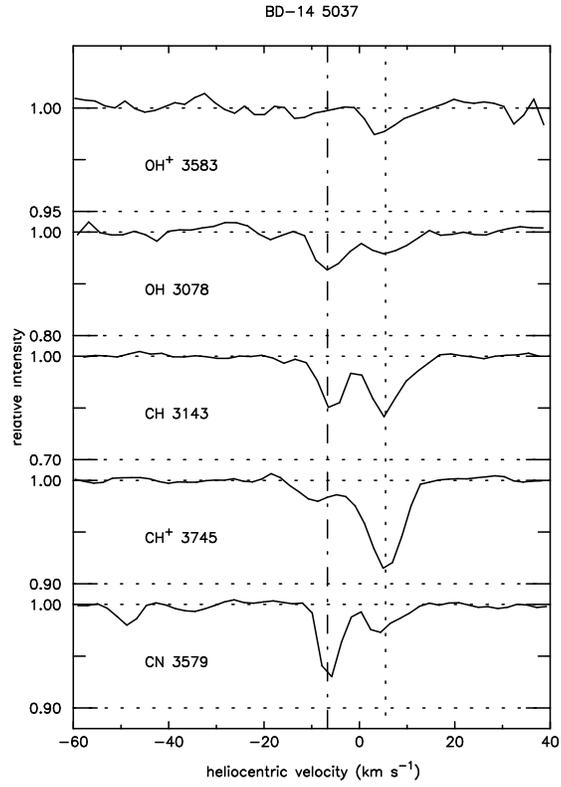}
\end{center}
\vspace{0.3in}
\caption{Absorption from OH$^+$, OH, CH, CH$^+$, and CN toward 
BD$-14^{\circ}5037$.  The wavelengths of each feature are indicated.  
The feature near $-$50 km s$^{-1}$ is the R(1) line of CN.  
Note that the vertical scales differ from panel to panel.  The dotted 
vertical line highlights the main OH$^+$/CH$^+$ component in other species.  
The dot-dashed line indicates the main OH/CN component.}
\end{figure}

\clearpage

\begin{figure}
\begin{center}
\includegraphics[scale=0.8]{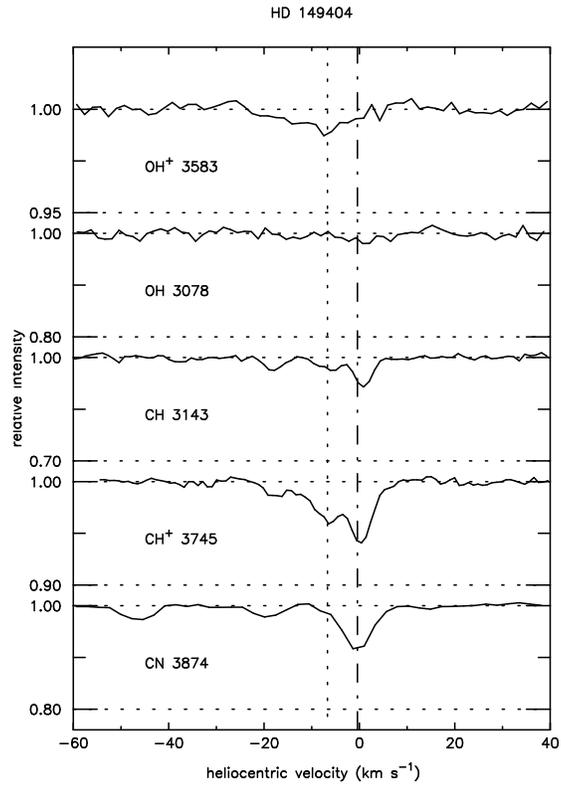}
\end{center}
\vspace{0.3in}
\caption{Same as Fig. 1 for absorption toward HD149404.  
CN $\lambda3579$ is very weak; the stronger line at 3874 \AA\
is shown.}
\end{figure}

\clearpage

\begin{figure}
\begin{center}
\includegraphics[scale=0.8]{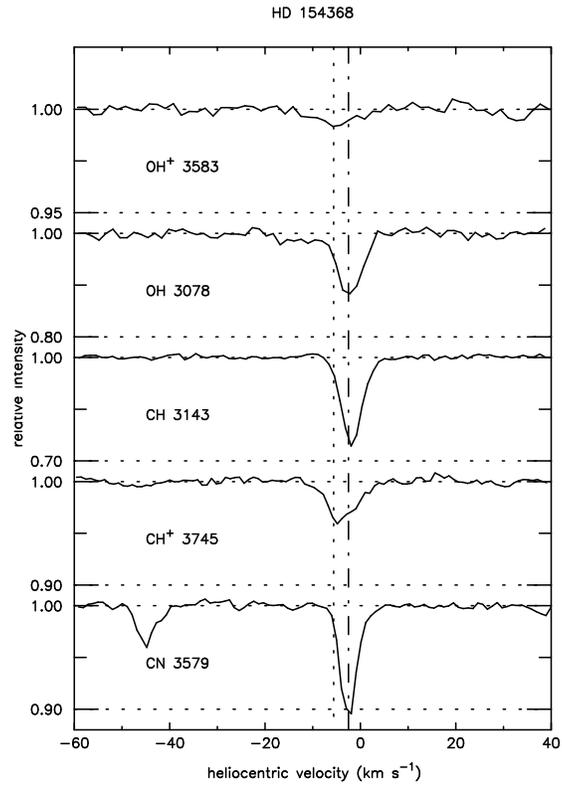}
\end{center}
\vspace{0.3in}
\caption{Same as Fig. 1 for absorption toward HD154368.}
\end{figure}

\clearpage

\begin{figure}
\begin{center}
\includegraphics[scale=0.8]{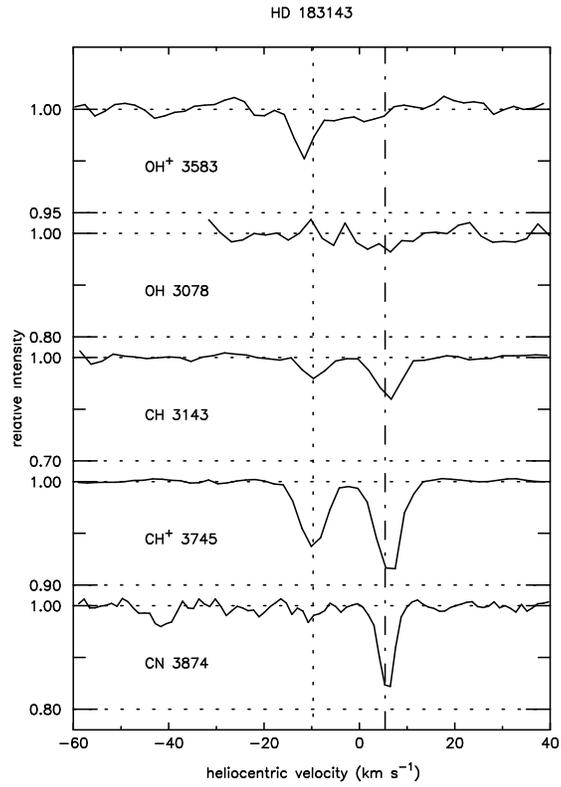}
\end{center}
\vspace{0.3in}
\caption{Same as Fig. 1 for absorption toward HD183143.  
CN $\lambda3579$ is very weak toward this star; 
CN $\lambda3874$ is shown instead.}
\end{figure}

\end{document}